\documentclass{article}

\AtBeginDocument{%
  }

\usepackage{multirow}
\usepackage{nicefrac}
\usepackage{microtype}
\usepackage{verbatim}
\usepackage{subcaption}
\usepackage{url}
\usepackage{pifont}
\usepackage{framed}
\usepackage{xcolor}
\usepackage{calc}
\usepackage{graphicx}
\usepackage{hyperref}

\definecolor{lavender_back}{HTML}{F3F0FA}
\definecolor{lavender_frame}{HTML}{3D305A}

\newcounter{customboxcount}

\newenvironment{custombox}[2][]{
    \refstepcounter{customboxcount}
    \par\medskip
    \setlength{\fboxrule}{0.4pt} 
    \setlength{\fboxsep}{8pt}  
    
    \MakeFramed{\advance\hsize-\width\FrameRestore}
    \noindent
    \textbf{\footnotesize\color{lavender_frame}%
      \texorpdfstring{Box \thecustomboxcount: #2}{Box \thecustomboxcount}}
    \par\medskip
    \hrule height 0.4pt
    \par\medskip
    \footnotesize
}{
    \endMakeFramed
}

\title{What You Prompt Is What You Get: Increasing Transparency Using Prompt Cards}

\author{
  Amandine M.~Caut \\ Uppsala University, Uppsala, Sweden
  \and
  Beimnet ~Zenebe \\ Addis Ababa University, Addis Ababa, Ethiopia
  \and
  Amy ~Rouillard \\ Stellenbosch University, Stellenbosch, South Africa
  \and
  David J. T. ~Sumpter \\ Uppsala University, Uppsala, Sweden
}

\date{} % remove if you want today's date

\begin{document}

\maketitle

\begin{abstract}
The rapid advancement and impressive capabilities of large language models (LLMs) have given rise to the field of prompt engineering, the practice of crafting inputs to guide LLMs toward high-quality, task-relevant outputs. A critical challenge facing the field is the lack of standardised prompt documentation and evaluation practices. Prompts can be long, complex and difficult to evaluate on subjective tasks. To address this challenge, we propose the use of prompt cards, structured summaries of prompt engineering practices inspired by the concept of model cards. Through prompt cards, the specific goals, considerations and steps taken during prompt engineering can be systematically documented and assessed. We present the prompt card approach and illustrate it on a specific task called "wordalisation", in which structured numerical data is transformed into text. We argue that a well-structured prompt card can enable better reproducibility, transparency, improve prompt methodology and give an effective alternative to benchmarking for judging the quality of generated texts. By systemically capturing underlying model details, prompt intent, contextualisation strategies, evaluation practices and ethical considerations, prompt cards make explicit the often implicit design decisions that shape system behaviour. Documenting these choices is important as prompting increasingly involves complex pipelines with multiple moving parts.
\end{abstract}


\section*{Introduction}

Prompt engineering is the art and practice of creating and optimising requests to guide large language models (LLMs) and other forms of generative artificial intelligence (AI) to produce desired outputs~\cite{mesko2023prompt,lazovsky2024art}. Although prompt engineering existed prior to the release of ChatGPT, its launch in November 2022 triggered a significant surge in the number of articles published in this field. A search on ArXiv for articles in the fields of computer science, electrical engineering and systems science using the keywords "prompt engineering", filtered by year, showed a growth from 2018 and 2019, when there were just 7 and 6 articles published, correspondingly, through a sharp increase from 83 in 2022, to 476, 1228 and 2086 articles in 2023, 2024 and 2025, respectively.

In the pursuit of prompts that improve, modify and tune the output of generative models, several new methods have arisen. Sahoo et al.~\cite{sahoo2025survey} classify the applications of prompt engineering into $12$ categories with a total of $41$ prompt techniques, including 'new task without data' (zero-shot and few-shot prompt techniques~\cite{brown2020language}) and 'reduce hallucinations' (Retrieval Augmented Generation~\cite{lewis2020retrieval}, ReAct~\cite{yao2023react}, CoVe~\cite{dhuliawala2023CoVe}, CoN~\cite{yu2024con} and Cok~\cite{li2024cok}). NeuroPrompts~\cite{rosenman2024neuroprompts}, for example, are used to optimise prompts for text-to-image generation by automatically enhancing a user's prompt. Moreover, prompt engineering is increasingly recognised as a valuable skill~\cite{oppenlaender2024prompt}.

One of several major concerns about LLM outputs is bias, as AI systems can inadvertently reproduce or amplify societal, cultural, or demographic prejudices leading to unfair or discriminatory outcomes~\cite{angwin2016machinebias, kiritchenko2018examining, Buolamwini2018GenderSI,zipperling2025s}. Another critical issue is hallucination, where AI generates information that is factually incorrect, misleading, or entirely fabricated, despite appearing plausible~\cite{maynez2020faithfulness, ji2023survey}. These phenomena can have serious consequences in high-stakes contexts, such as healthcare, legal decision-making, and scientific research, where erroneous or biased outputs may result in harm or the propagation of misinformation~\cite{singhal2023large, angwin2016machinebias}. Addressing these risks requires rigorous evaluation, transparency, and the development of safeguards to mitigate both bias and hallucination in AI systems.

The standard way to evaluate and compare the performance of models is benchmarking~\cite{McIntosh2025inadequacies,tichy1998should,wang2024benchmark}. Benchmarks have been proposed to measure labelling, paraphrasing and inference performance of prompting methods~\cite{wang2019glue}. However, a serious drawback of these benchmark tests is that, as Kaddour et al.~\cite{kaddour2023challenges} write, "slight modifications of the benchmark prompt or evaluation protocol can give drastically different results". Raji et al.~\cite{raji2021aiwideworldbenchmark} argue that focusing on benchmarks for every task is not always relevant, and not all tasks can or should be measured using standardised benchmarks. Their claim is supported by Bowman and Dahl~\cite{bowman2021what}, who write "we know of no simple test that would allow one to determine if a benchmark presents a valid measure of model ability" and Dehghani et al.~\cite{dehghani2021benchmark} who state that "in some cases, there is simply no standard benchmark or setup". 

Machine learning (ML) faces significant challenges regarding reproducibility, defined as the ability to replicate results under the same conditions as the original experiment~\cite{goodman2024repro, GUNDERSEN2022do}. Achieving reproducibility in AI is difficult and depends on several key factors:
\begin{itemize}
\item Access to the original dataset: Many datasets, especially in sensitive domains like healthcare, contain private information and are not publicly accessible.
\item Detailed methodological documentation: This includes all steps of the pipeline, such as data pre-processing and post-processing, which are often omitted or vaguely described.
\item Exact computational environment: Variations in hardware (e.g., CPU vs. GPU) or software (e.g., different library or framework versions) can lead to inconsistent results~\cite{beam2020challenges, semmelrock2023reproducibility}.
\item Complete model configuration: Training parameters (e.g., learning rate, batch size) are sometimes hidden, left at default values, or not reported at all.
\item Thorough result analysis: Final performance metrics and evaluation procedures must be transparently reported to enable meaningful comparisons~\cite{goodman2024repro}.
\end{itemize}
Even when all these conditions are met, inherent randomness in ML experiments, such as stochastic gradient descent or random weight initialisation, can cause variation in results, even under seemingly identical setups~\cite{beam2020challenges, semmelrock2023reproducibility}. As a result, obtaining identical outcomes across different runs or environments remains a significant challenge. 
Similarly, the same trained generative model can generate different outputs for identical inputs because sampling-based decoding introduces inherent randomness.

To tackle the ongoing challenges of reproducibility in AI, several initiatives have emerged, such as the ML Reproducibility Challenge from Princeton University~\cite{MLRC2025} and platforms like Hugging Face feature the "Trending Papers" page~\cite{pwc}, formerly known as Paper with Code, which evaluates the reproducibility of papers accepted at top machine learning conferences. These efforts aim to mitigate reproducibility issues by encouraging the validation and replication of published results. However, reproducing models remains challenging due to factors such as the use of proprietary or sensitive datasets, limited computational resources~\cite{semmelrock2023reproducibility}, and incomplete or non-functional code accompanying publications.

Recent work demonstrates that the behaviour of LLMs is highly sensitive to prompt phrasing, ordering, and contextual structure~\cite{perez2021true, lu2022fantastically, zhao2021calibrate,white2023prompt}. Unlike traditional classification and regression machine learning systems, where model behaviour is determined primarily by trained parameters, LLM outputs depend strongly on in-context examples, natural language instructions and retrieval-augmented information~\cite{brown2020language, akyurek2023learning, reynolds2021prompt}.

This sensitivity introduces reproducibility challenges: identical models can produce substantially different outputs when given slightly different prompts or contextual sequences~\cite{liu2023lost, berglund2024reversal}. As prompting, retrieval, and agentic orchestration now function as integral components of system behaviour, there is a growing need for prompt-level documentation. 

To improve the transparency and reproducibility of machine learning models, Mitchell et al.~\cite{mitchell2019model}  introduced the concept of a model card in 2019. A model card is a document that contains key information about a machine learning model. As a form of documentation~\cite{gebru2021datasheets, bender2018data,crisan2022interactive,crawford2021atlas}, model cards lie somewhere between a README file, which explains how to run the code, and detailed analyses of a method that are typically presented in academic publications. A model card differs from a README file in that it helps the reader to understand and reproduce the results, possibly modified for their own use case, as opposed to simply instructing a new user how to run the code. It includes all the model details and additional information such as intended use, ethical considerations and recommendations. By following a model card, users have a clear, step-by-step guide to reproduce the model and understand how it functions. This approach not only ensures reproducibility but also helps demystify the model's behaviour, making it easier for others to replicate, test, and build upon the work. Additionally, it provides crucial information about the model's limitations, enabling users to understand its boundaries and potential risks. It also enhances transparency and accountability by clearly pointing out weaknesses and outlining ethical considerations. By outlining both the model’s strengths and weaknesses, the model card fosters more informed and responsible use, helping users avoid misapplication or misinterpretation. Model cards have become widely used and are promoted by platforms such as Hugging Face\footnote{\url{https://huggingface.co/docs/hub/en/model-cards}}.

Model cards have traditionally been used to document trained models~\cite{mitchell2019model, gebru2021datasheets, bender2018data}, such as artificial neural networks. In this article, we adapt the model cards documentation framework proposed by Mitchell et al.~\cite{mitchell2019model} to prompt engineering with the aim of mitigating some of the issues raised above with respect to reproducibility, bias and accuracy. We propose the concept of a \textit{prompt card} and present a comprehensive description of its key elements, accompanied by an example based on wordalisation~\cite{wordalisation}. Our main contribution lies in solving the problem of transparency in prompt engineering by using a prompt card to document any process linked to prompt engineering. The prompt cards for each of our example applications are available on GitHub~\cite{wordalisation_PromptCard} and are displayed in the user interface of each application for easy reference\footnote{\url{https://wordalisations.streamlit.app}}.

\pagebreak
\section*{Prompt cards}

\begin{table}[h]
    \small
    \centering
    \begin{tabular}{c|cc}
    Sections     & Model Card & Prompt Card \\
    \hline
    Model Details    &  \ding{51} & \ding{51}\\
    Intended Use     & \ding{51} & \ding{51}\\
    Dataset & \ding{55} & \ding{51}\\
    Context & \ding{55} & \ding{51} \\
    Prompt architecture & \ding{55} & \ding{51} \\
    Factors     & \ding{51} & \ding{51}\\
    Metrics     & \ding{51} & \ding{55} \\
    Evaluation Data     &  \ding{51} & \ding{55} \\
    Training Data     & \ding{51} & \ding{55} \\
    Quantitative Analyses     &  \ding{51} & \ding{51}\\
    Data Security & \ding{55} & \ding{51} \\
    Ethical Considerations     & \ding{51} & \ding{51}\\
    Caveats \& Recommendations     & \ding{51} & \ding{51}\\
    \end{tabular}
    \caption{Comparison between sections included in a model card and a prompt card. A checkmark (\ding{51}) indicates a section that is relevant to a model or prompt card, and a cross (\ding{55}) indicates a section that is not applicable. The applicable model card sections are those suggested by Mitchell et. al.~\cite{mitchell2019model}.}
    \label{tab:modelvspromptcard}
\end{table}
%%%%%%%%%%%%%%%%%%%%%%%%

Our proposed categories for prompt cards, and their overlap with model cards, are summarised in Table \ref{tab:modelvspromptcard}. We discuss these categories, explaining how they can be used to describe prompts and highlight differences to model cards. To illustrate our approach, we document a prompt engineering approach called wordalisation~\cite{wordalisation}, using three example applications --- football scout, World Value Survey and personality test. The full prompt cards and the applications can be seen here~\cite{wordalisation_PromptCard}.

\subsection*{Model Details and Intended Use}

The \textit{Model Details} section for prompt cards should largely follow that of model cards~\cite{mitchell2019model}, containing the model’s date, version, license, and type, as well as details about its training algorithms, parameters, and other applied methods or features, where applicable. It should also reference supporting papers or resources, supply appropriate citation information and licensing terms, and specify a contact point for questions or feedback. For prompt engineering, it is important to reference the details of the LLM used during development. LLMs are frequently updated, and subtle differences between versions can significantly affect their outputs. For example, OpenAI and Google release new model versions every few months, which may include changes in system prompt, training data, architecture, or inference behaviour. By specifying the precise LLM used, future researchers can replicate results or consider potential variations caused by model updates. 

The \textit{Intended Use} section should also follow the recommendation in model cards~\cite{mitchell2019model}. Here, the specific tasks the model is designed to perform are described, the context in which it should be deployed, and any assumptions or limitations that users should consider. In the example application, see Box~\ref{box:intended-use}, we used the \textit{Intended Use} section to point out that our work is intended for educational purposes and professional football scouting %that using it to discuss data points, i.e. persons outside of the given dataset, 
is out of scope.

\begin{custombox}{Intended Use - Excerpt from the prompt card \textit{Intended Use} section for the football scout application.}
\label{box:intended-use}
The primary use case of this wordalisation is educational. It shows how to convert a Python Pandas Dataframe of football statistics about a player into a text about that player, which might be used by a scout or be of interest to a football fan. A secondary use case might be for users to understand more about the skills of male athletes playing in the Premier League during 2017-18. However, this version cannot be used for professional purposes, i.e. professional football scouting, partly because the data is out-of-date, but mainly because the functionality is limited. Professional use is thus out of scope. Use of the chat for queries not relating to the data at hand is also out of scope.
\end{custombox}

\subsection*{Dataset}

An important requirement for prompt cards is to carefully document any dataset used in the prompt. Such datasets can contain information related to specific input instances and may be independent of the prompt task. Task-dependent data—which might also be programmatically updated but usually remains fixed across input instances-should instead be discussed in \textit{Context}, see below. For wordalisation, the task is to produce accurate, fluent texts that discuss a specific candidate, football player, or country in the personality test, football scout, and World Value Survey applications, respectively. In these applications, the datasets contain the numerical descriptions of candidates, football players, and countries. Any data that provides task-specific context, for example few-shot example, is discussed in \textit{Context}. 

The dataset is a construct created by humans, inevitably encoding cultural norms, values and inherent biases~\cite{lee2019best}. Therefore, a thorough description of the dataset is essential to evaluating the LLM's output. Box~\ref{box:dataset} gives an example of a description of the dataset used in the personality application.  In Box~\ref{box:dataset}, we point the user to where the dataset can be accessed, give a broad overview of how the dataset is constructed, mention the number of candidates that responded, and clearly explain how each factor---Extraversion, Neuroticism, Agreeableness, Conscientiousness, and Openness---is computed. Further, a statistical model is applied, and each step and motivation is clearly documented.
\pagebreak
\begin{custombox}{Dataset - Excerpt from the prompt card \textit{Dataset} section for the personality test application.} 
\label{box:dataset}
The dataset used in this project was sourced from Kaggle's open dataset repository [www.kaggle.com/datasets/tunguz/big-five-personality-test], accessed July 2025. It consists of 1,015,342 questionnaire responses, collected online by Open Psychometrics. Respondents answered questions on a scale from 1 to 5, where: 1 = Strongly Disagree, 2 = Disagree, 3 = Neutral, 4 = Agree, 5 = Strongly Agree.\dots There are 50 questions divided into five categories: Extraversion, Neuroticism, Agreeableness, Conscientiousness, and Openness. Each question is associated with a weight of either +1 or -1, as determined by Open Psychometrics [https://openpsychometrics.org/printable/big-five-personality-test.pdf], accessed November 2025.\dots For each personality category, the score is calculated by summing the responses to the $10$ questions in that category.\dots [W]e compute the z-score for each category's final score, normalising the results to allow for comparisons across individuals and categories.
\end{custombox} 

\subsection*{Context}

When relevant context is provided in prompts, hallucinations by LLMs can be reduced~\cite{mei2025survey} and performance improved~\cite{shi2023llms, gao2024retrieval}. As mentioned in \textit{Dataset} section, the information which shapes this context may also convey societal norms or biases. Subsequently, the prompt card should document any contextual input they used to enrich the prompts. These can include system prompts and external information beyond the retrieved data, such as webpages, documents, curated question-and-answer (Q\&A) pairs, few-shot examples or other contextual signals~\cite{brown2020language, kojima2023llms}. Given these external dependencies, documentation should clearly specify all mechanisms used to augment or condition the model’s knowledge. This includes the origin and update schedule of curated context (e.g., Q\&A pairs, glossaries), the construction and selection criteria for few-shot examples, and full retrieval-pipeline details where retrieval-augmented generation (RAG)~\cite{lewis2020retrieval} is employed (e.g., embedding model versions, vector dimensionality, chunking strategy, indexing method, retrieval thresholds). RAG can encompass text~\cite{guu2020realm}, images~\cite{feng2024promptmagician, wang2023diffusion}, and numerical data~\cite{xue2023promptcast}, either individually or in combination, as in multimodal prompting approaches~\cite{nooralahzadeh2025multimodal, son2025advancing, sheng2025mqrld}, and this should be clearly documented in \textit{Context}.

Additional knowledge-injection methods, such as rule-based templates or structured task schemas, should also be described, as they introduce systematic constraints on model behaviour. Together, these components ensure that contextual augmentation remains reliable, current, and aligned with the model’s intended use~\cite{mitchell2019model}. In the examples below, the prompt card documents the user instructions (Box~\ref{box:user-instructions}), the process used to construct Q\&A pairs used as part of the prompt (Box \ref{box:QandA}) and describes the user-assistant pairs used for few-shot prompting (Box \ref{box:few-shot}). These examples guide the LLM in adopting a consistent tone, provide relevant information and clarify how it should structure its answers. 

Further, for wordalisation, a normative model is used to transform the z-scores, discussed in \textit{Dataset}, into textual descriptions (See Box~\ref{box:normative-model}). Clearly documenting these steps is important as we are injecting our norms and values into the data: for instance, in the context of the football scout example, we might describe a player as "below average" performer, based on a specific metric, but if we were to compare that same professional player to the general population, the term "below average" would no longer be meaningful. This step is also judgmental in the sense that we are passing judgment on a player's performance using statistics, which they may see as unfair or hurtful. Similar issues apply to personality tests and the World Values Survey examples. For the World Value Survey, norms vary across countries and evolve over time. For personality tests, critics argue that the Big Five personality traits~\cite{john2008paradigm} may not be universally applicable, as cultural differences can affect how traits are perceived and measured~\cite{raad2010only,triandis2002cultural}.

\begin{custombox}{Context - Excerpt from the prompt card \textit{Context} section documenting the user instructions given for the personality test application.}
\label{box:user-instructions}
The following "role" is given to the LLM: "You are a recruiter who provides concise, data-driven summaries of candidates. You rely solely on the provided information to deliver clear and to-the-point evaluations." These instructions are brief and to the point, and ensure that the responses are relevant to the target audience.
\end{custombox}

\begin{custombox}{Context - Excerpt from the prompt card \textit{Context} section discussing the context provided by Q\&A pairs for the personality test application.}
\label{box:QandA}

We created Q\&A pairs to represent each personality trait in the personality test dataset used to generate wordalisation to describe individuals from the dataset. To ensure these pairs accurately reflect each trait, we used descriptions from Wikipedia to contextualise the measurements in the data. Each trait—Extraversion, Conscientiousness, Openness, Agreeableness, and Neuroticism—is represented through multiple question-answer pairs derived from Wikipedia's explanations. The goal is to identify the questions that best capture the essence of each trait through their corresponding answers. Using the Wikipedia description of 'Openness': "Openness to experience is a general appreciation for art, emotion, adventure, unusual ideas, imagination, curiosity, and variety of experience. People who are open to experience are intellectually curious, open to emotion, sensitive to beauty, and willing to try new things. [...]", [we construct the following user-assistance pair]:

\begin{tabular}{p{5.5cm} | p{\linewidth-5.5cm-4\tabcolsep-20pt-1.2pt}}
\multicolumn{1}{c}{\textbf{User}} & 
\multicolumn{1}{c}{\textbf{Assistant}} \\
\hline
\dots & \dots \\
How might people with high openness to experience be perceived? & Open people can be perceived as unpredictable or lacking focus, and more likely to engage in risky behaviour or drug-taking. Moreover, individuals with high openness are said to pursue self-actualisation specifically by seeking out intense, euphoric experiences. \\
\dots & \dots

\end{tabular}

We repeat this process for the four remaining Big Five personality traits. These examples enable the system to acquire insights into each personality trait, equipping the chat with professional expertise.
\end{custombox}

\begin{custombox}{Context - Excerpt from the prompt card \textit{Context} section for the personality test application discussing the application of a normative model to produce synthetic text describing a candidate.} 
\label{box:normative-model}
Using the z-scores, we generate a personalised paragraph for each individual. Each paragraph contains at least two phrases reflecting each of the five personality traits. The phrasing is determined by the z-score, where one phrase is influenced by the direction of the z-score (positive or negative) and the other incorporates an adjective based on its magnitude. The positive or negative sign of the z-score dictates the tone, while the magnitude of the z-score is used to assign an adjective that characterises the trait. For example, for Extraversion, an individual will be described as outgoing and energetic when the z-score is positive, accompanied by the phrase, "The candidate tends to be more social." For a negative z-score, extraversion will be characterised as solitary and reserved using the phrase, "The candidate tends to be less social."
\end{custombox}

\begin{custombox}{Context - Excerpt from the prompt card \textit{Context} section discussing the context provided by few-shot prompting for the personality test application.}
\label{box:few-shot}

Using few-shot prompting, we aim to demonstrate to the LLM how to answer. For this purpose, we create user-assistance pairs that demonstrate to the LLM how to convert the synthetic text describing a candidate into a "wordalistation". In this application, we use four examples, with one shown below. These examples are used to emphasise that we wish to highlight positive and negative aspects of the person's personality, and ignore aspects which are average or typical.

\begin{tabular}{p{5.5cm} | p{\linewidth-5.5cm-4\tabcolsep-20pt-1.2pt}}
\multicolumn{1}{c}{\textbf{User}} & 
\multicolumn{1}{c}{\textbf{Assistant}} \\
\hline
The candidate is very outgoing and energetic. The candidate tends to be more social. In particular, they said that they start conversations. The candidate is quite sensitive and nervous. The candidate tends to feel more negative emotions and anxiety. The candidate is quite friendly and compassionate. The candidate tends to be more cooperative, polite, kind and friendly. The candidate is very efficient and organised. The candidate tends to be more careful or diligent. In particular, they said that they pay attention to details. The candidate is relatively consistent and cautious. The candidate tends to be less open to new ideas and experiences. 
&
The candidate appears to be an outgoing and energetic individual who enjoys social interaction and readily engages others in conversation. They demonstrate strong organisational skills, attention to detail, and a cooperative nature, though their sensitivity may at times make them prone to stress or self-doubt. They are likely to thrive in structured, collaborative environments that value precision, clear expectations, and positive interpersonal dynamics
\end{tabular}

\end{custombox}

\subsection*{Prompt architecture}
Model cards often include a schematic representation of the model architecture, enabling readers to understand it immediately. Similarly, prompt cards should provide a clear breakdown of the entire prompt, enabling transparency, clarity and easier debugging. Such a prompt schematic should describe the prompt structure, delineate the different components of the prompt, their roles and how the prompt is assembled. Documenting the prompt structure in this manner allows others to understand its design, replicate results, and adapt it safely for new applications~\cite{openai2025gpt52prompting}. Figure \ref{box:architecture} gives an overview of the prompt architecture for wordalisation~\cite{wordalisation} prompt methodology.

\begin{custombox}{Prompt architecture - Excerpt from the prompt card \textit{Prompt architecture} section}
\label{box:architecture}
The figure below summarises the wordalisation prompting methodology. The "Tell it who is it", "Tell it what it knows", "Tell it what data to use" and "Tell it how to answer" steps are discussed in the \textit{Context} section and correspond to task-specific information and instructions. The raw data, data preprocessing and statistical model (z-scores) are documented in section \textit{Dataset}.

\centering
\includegraphics[width=0.6\linewidth]{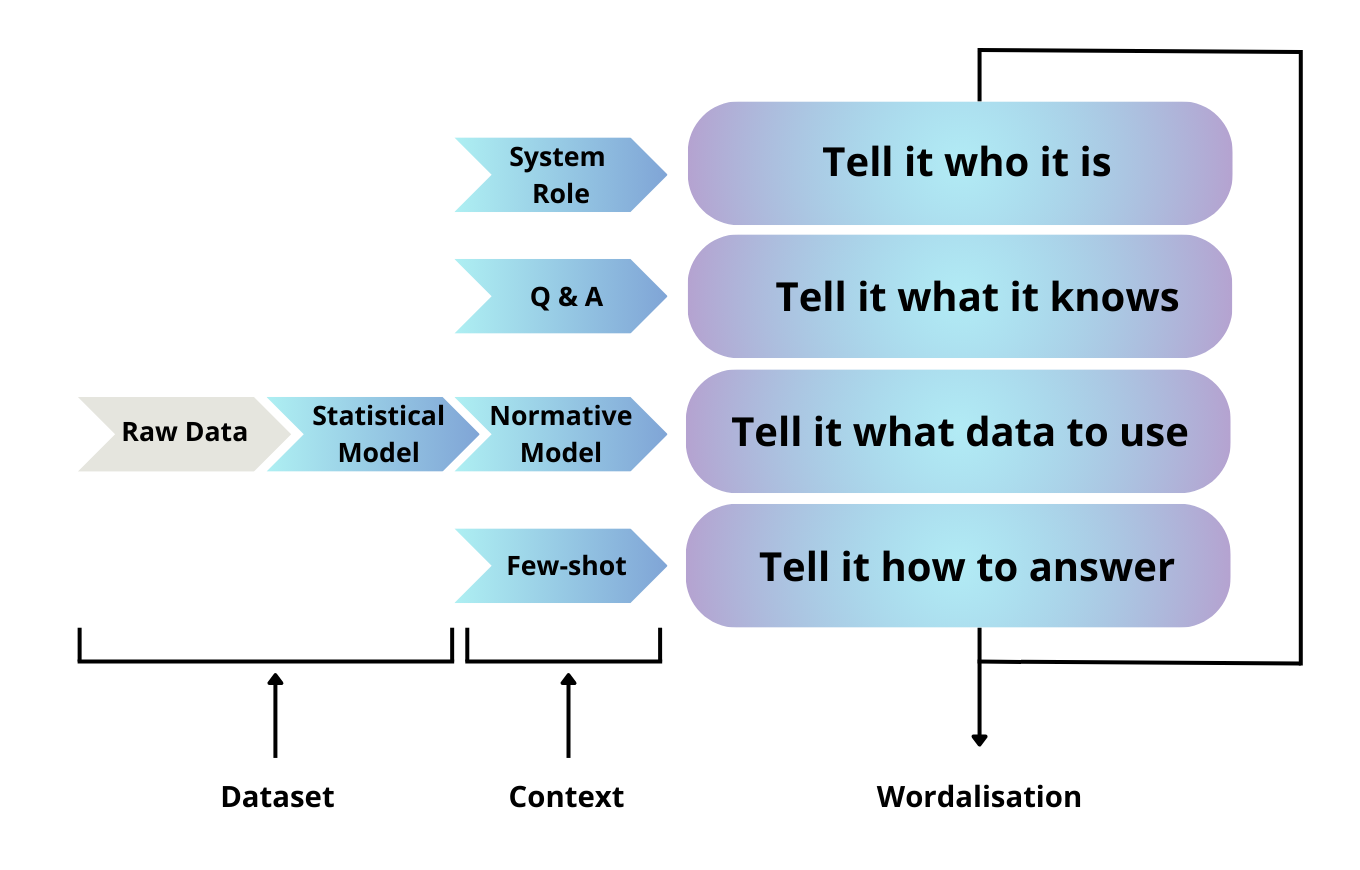}

Figure 1: Prompt architecture of the wordalisation process

\end{custombox}

\subsection*{Factors}

The \textit{Factors} section in the model card summarises performance across key dimensions, such as demographic groups, instrumentation types, and environmental conditions. Documenting this information is important since it reveals potential biases, highlights how the model behaves under different conditions, and informs the user about possible issues in the dataset. 

Similarly, the prompt card should also document any important factors related to the dataset used. For example, for the World Values Survey, see Box~\ref{box:factors-country}, we point out that the dataset used covers 66 countries and took place from 2017 to 2022. We emphasise that outputs related to non-included countries lack empirical support and show that survey participants represent varying, often small, samples of the population. Therefore, the values and statements in the wordalisation should not be viewed as representative of entire populations. The football scout application uses a publicly available dataset, see Box~\ref{box:factors-player}. However, the recognisability of the individuals involved brings its own ethical challenges. Further, the dataset is limited to a specific demographic: male professional football players. 

\begin{custombox}{Factors - Excerpt from the prompt card \textit{Factors} section for the World Value Survey application.}
\label{box:factors-country}
The World Value Survey data and derived factors, discussed in section \textit{Dataset}, relate to 66 countries that took part in the WVS "wave 7" 2017-2022 survey. We would like to state that any reports or chats about countries not included in the survey are not guaranteed to hold any merit. We also note that the participants of the "wave 7" survey constitute only a small sample of the population of each country, see Figure [1]\dots. Therefore, the values and statements presented in the app should not be considered representative of the entire population of any given country.

\centering
\includegraphics[width=0.6\linewidth]{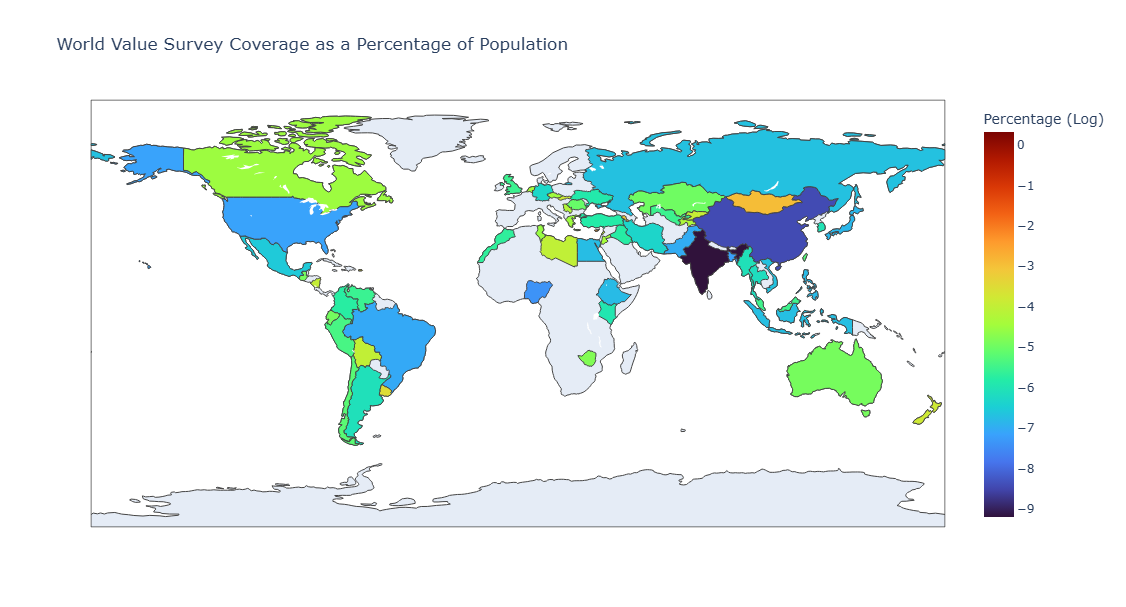}

Figure~[1]: World Value Survey data collection distribution as a percentage of the country's population. The populations were taken from the [United Nations Department of Economic and Social Affairs](https://population.un.org/wpp/Download/Standard/CSV/) data on world population prospects in 2017 (the first year of "wave 7"). The colour scale indicates the log of the percentage of the population of each country that participated.
\end{custombox}

\begin{custombox}{Factors - Excerpt from the prompt card \textit{Factors} section for the football scout application.}
\label{box:factors-player}

The football scout wordalisation is applied to a very specific demographic group, namely male professional football players. This version is only for use within that group and thus excludes female athletes. The ethnicity and social background of the players are not documented, but can be (anecdotally) considered to be more diverse than that of the English population as a whole (the Premier League is played in England). Players come from all over the world to play in the Premier League. The dataset was chosen because of the availability of a public dataset and because of the fact that the players will be recognisable names for many users.
\end{custombox}

\subsection*{Metrics, Evaluation Data and Training Data}

The section \textit{Metrics}, \textit{Evaluation Data}, and \textit{Training Data} in model cards are not relevant for prompt cards. Prompts typically do not involve training a model from scratch or optimising decision thresholds. Instead, they rely on pre-trained large language models whose behaviour may vary depending on context, phrasing, or randomness in generation. While one can evaluate prompt performance using human judgment or automated scoring of outputs (see below), the standardised quantitative metrics (such as accuracy or AUC) used for model cards cannot be straightforwardly applied to prompts, making reproducibility and evaluation challenging.

\subsection*{Quantitative Analysis}

Quantitative analysis in the context of prompt cards serves a different purpose from that in model cards. Rather than providing a comprehensive, benchmark-style characterisation of system performance, quantitative methods are used to investigate specific, well-defined hypotheses about prompt behaviour. For example, whether generated text accurately conveys provided data or whether particular prompt components materially affect output quality. In this sense, quantitative analysis supports the evaluation of specific aspects of the prompt's effectiveness, rather than offering a complete assessment of its overall quality.

For tasks with established benchmarks—such as summarisation~\cite{zhang2023benchmarking}, question-answering~\cite{rajpurkar2016squad}, or sentiment analysis~\cite{wang2019glue}—these resources can sometimes be repurposed to evaluate the outputs produced by prompts. However, for the majority of real-world applications, no suitable benchmarks exist. Indeed, one motivation for constructing a prompt card is to document prompt behaviour in the absence of standardised benchmarks, on a per-application and per-use-case basis.

Model cards typically include a \textit{Quantitative Analysis} section reporting unitary and intersectional results across controlled testing datasets. Such evaluations are not always directly transferable to the context of prompt engineering. Prompts do not constitute trained artefacts with stable, reproducible behaviour. Instead, their outputs depend on the underlying LLM, prompt phrasing, contextual information, and sources of stochastic variation. As a result, it is generally impractical and often misleading to aim for standardised quantitative metrics that are comparable to those used in model cards. Quantitative analysis for prompts should therefore be targeted, contextual, and hypothesis-driven.

As illustrated in Box~\ref{box:evaluation}, we use quantitative evaluation to test a specific hypothesis, namely that the wordalisation prompt faithfully conveys the information contained in the underlying synthetic text. To do so, we employ an LLM-as-judge approach~\cite{gu2025survey}, in which an LLM is tasked with reconstructing structured information from generated wordalisations and comparing this reconstruction to the original data. This enables a measurable notion of faithfulness, operationalised as reconstruction accuracy. LLM-as-judge methods offer scalable and internally consistent evaluation by applying the same natural-language criteria across large samples, thereby reducing some sources of variability present in human ratings~\cite{Croxford2025evaluating}. At the same time, such methods may inherit model biases, produce opaque or hallucinated reasoning, and be highly sensitive to prompt design~\cite{Croxford2025evaluating, shi2025judging}. These limitations mean that the results should be interpreted as evidence with respect to a narrow claim, rather than as a definitive assessment of quality.

Human evaluation can also provide a complementary perspective. Human reviewers are better suited to assessing subjective and context-dependent qualities such as fluency, tone, engagement, cultural appropriateness, and practical usefulness, and can offer richer qualitative feedback grounded in ethical and factual judgment. However, human evaluation is costly, slow to scale, and prone to inconsistency and bias in the absence of substantial oversight~\cite{Croxford2025evaluating, lee2019best, amidei2018rethinking}. Moreover, as noted by Belz et al.~\cite{belz2023missing}, the majority of human evaluations in NLP lack repeatability or reproducibility, which limits their utility as standalone evidence. If human evaluation is used, this section of the prompt card should document all relevant aspects of the evaluation design, including the evaluation task, rater demographics and expertise, annotation guidelines, sample size, aggregation method, and any measures taken to assess or improve inter-rater agreement.

Quantitative evaluation, whether performed by an LLM judge or a human rater, plays an important role in prompt development as it is particularly useful for probing narrowly defined properties, such as accuracy or faithfulness. It can also help surface issues related to usefulness, limitations, and potential failure modes. In contrast to typical machine learning evaluation, quantitative analysis in the prompt card should be understood as just one component of the overall discussion presented in the prompt card. Text generation is a complex task involving multiple facets, including fluency, engagement, accuracy, faithfulness, and relevance, that cannot be fully captured by any single metric. By situating targeted quantitative findings alongside information about intended use, retrieved and contextual data, prompt structure, influencing factors, and data security and ethical considerations, the prompt card provides users with a more complete and nuanced picture of the prompt’s strengths and weaknesses, supporting informed and context-sensitive deployment.

\begin{custombox}{Quantitative analysis - Excerpt from the prompt card \textit{Quantitative analysis} section for the World Value Survey application}
\label{box:evaluation}
For each data point in the dataset, we generate a sample of wordalisations for evaluation. For comparison, we also generated wordalisations using a version of the prompt that did not contain the relevant synthetic text generated by the normative model. To discourage the LLM from declining to respond, we added the sentence `If no data is provided, answer anyway, using your prior statistical knowledge', and modified one of the in-context learning examples by removing synthetic text from the user prompt while leaving the wordalisation (response) unchanged. For consistency, the same prompt template was used both when data was and was not provided. To take into account random variations in the wordalisations due to the stochastic nature of the LLM, we passed each prompt to the LLM multiple times to generate a set of wordalisations for evaluation. 

To assess the accuracy of the text generated using the wordalisation prompt, we used an LLM to reconstruct the original information provided in the synthetic text from the wordalisation. The prompt for this task takes the form of a multiple-choice question (MCQ) for each factor (traditional vs secular values, survival vs self-expression values, neutrality, fairness, scepticism and social tranquillity). We then compare the `true' class according to the normative model with the reconstructed data class to measure how faithfully the wordalisation represents the original data. Figure [1] shows the results of this analysis. 

\centering
\includegraphics[width=0.4\linewidth]{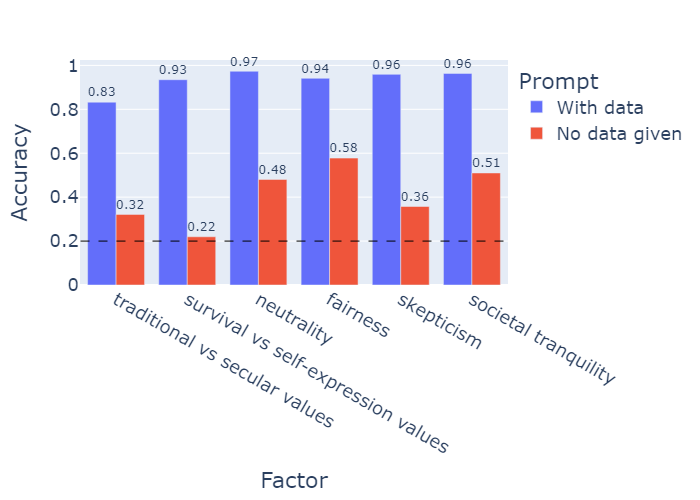}

Figure [1]: Comparison of the class labels generated by the normative model with classes reconstructed from the wordalisations. Multiple wordalisations were generated for each data point, so that at least $10$ valid reconstructions per data point were found, and the mean accuracy is taken over all wordalisations. The accuracy is compared for two different prompts, one in which data in the form of synthetic texts was given (purple) and in the other, the data was omitted (red). The dashed line indicates the expected accuracy if the class labels were randomly chosen according to a uniform probability distribution ($\frac{1}{5}$). 

\dots This approach has some weaknesses, including relying on the LLM to generate accurate reconstructions and to adhere to the MCQ options. In cases where the wordalisations are more formulaic, this type of evaluation can work well, as was the case here. However, in cases such as where irrelevant or uninteresting information is omitted, for example, that a given country has average attitudes toward `fairness' compared to other countries in the same survey, reconstruction is more nuanced.
\end{custombox}

\subsection*{Data security}

One major issue with LLMs is the security and vulnerabilities~\cite{he2025the, wu2024new, jiao2025navigating}, such that the potential leakage of personal and sensitive details provided in the prompt~\cite{WU2024unveiling}. When using online third-party LLM services, information is transmitted to the service provider, such as Google, OpenAI or Anthropic, which hosts the model on their servers or cloud infrastructure. Those details can be saved, logged, or temporarily stored by the service provider. In some cases, this data might be inadvertently or intentionally included in a future training dataset, leading to the risk of the data being leaked or exposed through subsequent model outputs or security breaches. The ability of online LLMs to access the internet introduces another significant security and privacy concern, as it expands the surface area for potential data exposure and interaction with external, unverified sources.

To mitigate these risks and improve transparency, we recommend adding the following security-relevant fields to the Prompt Card:
\begin{itemize}
    \item Internet use: Indicate if the prompt requires the LLM to access the internet or external APIs.
    \item Offline adaptability: Note if the prompt can be adapted for a privacy-focused, offline, or locally hosted LLM version.
    \item Security assessment: List any specific security issues that need to be taken into account when using the prompt (e.g., handling sensitive customer data, private health information (PHI), processing proprietary information).
\end{itemize}

Medical, financial and other personal-data use cases of LLMs provide concrete examples of the considerations that arise when sensitive user data is shared with third-party services  
~\cite{yao2024llm, duan2024privacy}. For instance, patient medical records are highly sensitive, and transmitting names of patients and even potentially identifying factors such as age and location or certain raw metrics to an external LLM may violate clinical data-handling regulations. 
Interaction with an LLM is not, however, a binary choice between sharing full data or none at all. There exist different levels of privacy risk, depending on how much information is disclosed and how directly it relates to an identifiable individual~\cite{wirth2021data, im2024privacy}. At one extreme, prompts may include raw values or complete records, which represent the highest risk to data privacy. At intermediate levels, sensitive data can be locally processed, transformed, or abstracted before being included in a prompt. For example, rather than sending full medical records, it is possible to first compute clinically relevant summaries or model-derived interpretations and then pass only a derived explanation, such as "the patient's glucose levels place them at a higher cardiovascular risk" in the prompt. Such preprocessing has the power to shift the prompt content from containing sensitive, identifiable and raw metrics to interpretive descriptions~\cite{im2024privacy}. 
Prompt authors should detail considerations made on how information can be passed to the LLMs in a way that avoids personally identifiable or high-risk attributes. Box \ref{box:security-personality} shows how, in the personality test example, the prompt containing the individual data is written in such a way that it avoids providing the test score directly. 

\begin{custombox}{Data Security - Excerpt from the prompt card for the personality test application, showing an example of an anonymised synthetic text.}
\label{box:security-personality}
The candidate is very outgoing and energetic. The candidate tends to be more social. In particular, they said that they start conversations. The candidate is quite sensitive and nervous. The candidate tends to feel more negative emotions and anxiety. The candidate is quite friendly and compassionate. The candidate tends to be more cooperative, polite, kind and friendly. The candidate is very efficient and organised. The candidate tends to be more careful or diligent. In particular, they said that they pay attention to details. The candidate is relatively consistent and cautious. The candidate tends to be less open to new ideas and experiences.
\end{custombox}

\subsection*{Ethical considerations}

Ethical concerns surrounding large language models are multiple and interconnected. Core issues include bias and fairness, privacy and data security, and the risks of misinformation and disinformation, all of which shape how these systems affect society~\cite{jiao2025navigating, gallegos2024bias, yao2024llm}. Prompt-based systems may encode normative assumptions about categories, values, or behaviours, thereby reinforcing societal stereotypes or marginalising alternative perspectives. 
At a societal level, LLMs can reinforce existing prejudices, contribute to discrimination, and harm marginalised groups, while also raising concerns about unequal access and widening social inequality~\cite{weidinger2021ethical,klein2024data, bender2021dangers}. 
Safety and security risks are already being taken into consideration in the section \textit{Data security}. 
Questions of integrity are central as well, particularly around transparency and accountability in model design and deployment, intellectual property and plagiarism, and the environmental impact of training and operating large models~\cite{li2025making}. 
Information is vital today because it shapes decisions and influences societies. However, when language models leak or infer sensitive information, it can lead to privacy violations, misinformation, discrimination, and loss of trust in digital systems~\cite{weidinger2021ethical}. LLMs raise concerns related to the trustworthy generation of false, misleading, or oversimplified information, which may be especially harmful when such systems are used in sensitive domains, such as healthcare, legal reasoning or public policy~\cite{han2024medical,bommasani2022foundation}. 
Censorship introduces additional ethical concerns~\cite{glukhov2023llm}; while content moderation can prevent harmful outputs, it also raises concerns about subjectivity, the potential infringement on free speech, and the lack of transparency in moderation policies. 
Finally, intellectual property and plagiarism pose significant legal and ethical challenges, including the use of copyrighted material in training data, the protection of intellectual property rights, the boundaries of fair use, and questions of liability for generated content  ~\cite{wei2025interrogating}.

As in the model card, the prompt card should document ethical considerations associated with the use of prompt-based systems. These include potential biases, representational harm, and impacts on individuals or groups affected by the data or model ~\cite{gallegos2024bias} or any other sensitive information mentioned above.

As a concrete example, when analysing football players, we acknowledge that their performance is being evaluated in words by an automated system that does not understand broader factors in their lives. As a further illustration, each social factor in the World Values Survey application is calculated from a set of related questions, which are assumed to be correlated with a broader social value. For example, a series of questions about whether certain criminal activities are ever justifiable is labelled as a measure of fairness in society. In light of these methodological assumptions and limitations, we also highlight existing criticisms of the analytical framework applied~\cite{haller2002theory, aleman2016value}, and strongly encourage domain experts to consider these critiques when interpreting or leveraging this data. The personality test faces a similar critique, as some statements may exhibit bias when interpreted across different cultures~\cite{van2004bias}. When using personality test data (Box~\ref{box:ethical}), we therefore acknowledge its cultural bias and limited universality.

More broadly, the ethical considerations section of a prompt card should make explicit how prompt-based systems might affect users, particularly when outputs are used to inform judgment or decision-making. It should detail known failure modes, uncertainty characteristics, and contexts in which outputs should not be treated as authoritative. By documenting known biases and limitations, prompt cards support more responsible interpretation, discourage inappropriate use, and align prompt-based system development with emerging expectations for transparency and accountability in AI systems.

\begin{custombox}{Ethical considerations - Excerpt from the prompt card \textit{Ethical Considerations} section for the personality test application}
\label{box:ethical}
Several ethical challenges have arisen in this project. First, there are no professional psychologists or psychiatrists involved in interpreting the results. The interpretation is done by a statistical method as mentioned above. Second, the Big Five personality test has several criticisms, such as being overly simplistic~\cite{abood2019big} and being subject to cultural bias. Indeed, according to~\cite{deraad2010three, triandis2002cultural, costa2001gender}, the personality is developed in an environmental and cultural context. The test can not be applied universally~\cite{abood2019big, deraad2010three}, and should have more traits~\cite{abood2019big}. Third, the test can be easily manipulated by individuals seeking to present a more favourable personality. Last but not least, we must ask ourselves whether we truly want to automate the hiring process and risk losing the human element. Without this, we may end up with companies that consistently recruit the same type of profile~\cite{Bach2005Managing}. When a company automates its recruitment process, it often starts by defining the type of candidate already present within the organisation, which can inadvertently create a biased dataset~\cite{bogen2018help, bogen2019all}. This leads to the selection of standardised candidates who closely resemble existing employees. If the candidates mirror the profiles of those already in the company, how can we expect to bring in fresh perspectives and innovative ideas? Careful consideration should be made if applying these methods to data collected on athletes in environments where this type of scrutiny isn't the norm.
\end{custombox}
 
\subsection*{Caveats and recommendations}
This section should address additional considerations that were not covered in the preceding sections.
For example, an overview of the known limitations, uncertainties, or potential pitfalls associated with the method, tool, or findings presented. It should highlight any assumptions made, conditions under which the results may not hold, and areas where performance may degrade. Additionally, this section should offer practical guidance for users or researchers on how to responsibly apply the method, avoid misinterpretation, and improve outcomes. Recommendations may include best practices, suggestions for future improvements, or complementary techniques to mitigate the identified caveats. The goal is to promote transparency, responsible usage, and continuous refinement.

\section*{Discussion}

Prompt engineering is widely used today, from casual users interacting with ChatGPT to professionals integrating large language models into their applications. LLMs have the potential to make all types of data more accessible and comprehensible to a wider audience. Evaluating prompt engineering could be viewed as challenging, as it is inherently subjective~\cite{tong2021inapplicability, ieee_chatgpt}. We have embraced this reality by prioritising transparency and honesty, akin to practices in visualisation~\cite{mintzer-sweeney_muth_2024,zhao2023stories,cairo2016truthful}. Rather than emphasising numerical evaluations, we advocate for providing thoughtful documentation through prompt cards as a more holistic approach to communicating model capabilities. We now argue that benchmarking and evaluation should support this process rather than be the sole target. 

We can compare the process of best practices with journalism: journalists rarely focus on benchmarking their work, but aim to maintain high standards. Data journalism addresses the challenge of how to filter large volumes of data, interpret it, visualise it and present it in a compelling way to tell meaningful stories. The process typically involves collecting data from sources such as public databases or through web scraping, cleaning and organising the information, analysing trends and patterns, and finally visualising the results using charts, maps, or interactive tools. By transforming raw data into clear and accessible narratives, data journalism provides audiences with a deeper, evidence-based understanding of complex issues in the world around them. Benchmarking everyday data in data journalism isn’t usually possible because data lacks structure or historical parallels.

Many tasks for prompting, like the wordalisation task we focus on here, also do not fit into benchmarking frameworks. To see why this is the case, consider how
Wang et al.~\cite{wang2019glue} categorise NLP benchmarks into three types: single-sentence tasks, similarity and paraphrase tasks, and inference tasks.
In the case of single-sentence tasks, the benchmarks CoLA~\cite{CoLA} and SST-2~\cite{SST-2} are designed for labelling, such as sentence acceptability and sentiment prediction, and differ significantly from the aim of wordalising data. Moreover, similarity and paraphrase tasks, where the goal is to assess whether sentence pairs or question-answer pairs are similar or semantically equivalent~\cite{MRPC,QQP,STS-B}, also do not align with the needs. That said, in the "tell it how to answer" step, where we rephrase and refine the generated text using models like ChatGPT\footnote{\url{https://github.com/openai/simple-evals?tab=readme-ov-file\#benchmark-results}} and Gemini\footnote{ \url{https://deepmind.google/technologies/gemini/}}, we could consider relevant benchmark tests from those models themselves. For inference tasks, the benchmarks focus on evaluating prediction performance from a given text. For example, the Multi-Genre Natural Language Inference Corpus (MNLI)~\cite{MNLI} and the Recognising Textual Entailment (RTE)~\cite{RTE1,RTE2,RTE3,RTE5} datasets evaluate whether a given hypothesis is confirmed, contradicted, or neutral based on the provided context. Similarly, the Stanford Question Answering Dataset (QNLI)~\cite{QNLI} focuses on classifying sentence pairs by determining whether a question is supported by its corresponding sentence. Finally, the Winograd Schema Challenge (WNLI)~\cite{winograd}, on the other hand, assesses a model's ability to correctly resolve pronouns in ambiguous contexts. While these benchmarks are valuable for a variety of NLP tasks, they do not align with the objectives of our prompt engineering approach.

We see the wide variety of potential LLM applications as an illustration of Bowman and Dahl's assertion that "benchmarking for NLU (Natural Language Understanding) is broken". 
While benchmarks have proven valuable in advancing scientific research in certain domains, their application to LLMs has largely fallen short. Unlike the structured and standardised benchmarking systems found in regulated industries, it is not the case in AI domain~\cite{McIntosh2025inadequacies}. This absence has led to a proliferation of individually designed benchmarks~\cite{raji2021aiwideworldbenchmark,McIntosh2025inadequacies}. Considering each prompt is designed to address a specific problem within a particular domain, there is no universal benchmark to evaluate every individual prompt. With the growth of prompt engineering techniques, the idea of benchmarking everything uniformly has become unrealistic~\cite{raji2021aiwideworldbenchmark}. McIntosh et al.~\cite{McIntosh2025inadequacies} study compares 23 state-of-the-art benchmarks applied to LLM, they point out "significant inadequacies across technological, processual, and human dynamics" that reduce their effectiveness and credibility. They also highlight the lack of standardised frameworks in AI benchmarking.
Another arising issue is benchmark data contamination, i.e. when elements of the evaluation benchmarks are inadvertently included in the training data of the LLM itself~\cite{xu2024contamination}. How is it possible to evaluate a language model in this case? Since most of the LLMs are trained on internet data, there's a risk that the LLM train itself on the benchmark test sets~\cite{brown2020language, xu2024contamination}. This leads to contamination, where the model is effectively tested on data it has already seen, resulting in unreliable performance scores and misleading evaluations.

We believe that research should shift its focus away from developing yet another benchmark and instead prioritise exploring novel ideas to address both current and emerging challenges.
While it is neither practical nor necessary to create benchmarks for every possible task~\cite{raji2021ai}, growing concerns about the validity of many existing AI benchmarks have led experts to advocate for fewer, but more rigorous and comprehensive benchmarks~\cite{Ott2022mapping, Geirhos2020shortcut}. As a result, developing methods to enhance the transparency of AI tools, as we have done here, has become increasingly important. 

Recent regulatory efforts, for example, the EU's AI Act, the US's National Artificial Intelligence Initiative Act and South Korea's AI Basic Act, demonstrate growing legislative and societal concerns over AI risks. Yet, as Haataja \& Bryson argue, even strong regulatory frameworks may fail to enforce safe practices if AI systems remain vague and un-auditable~\cite{haataja2023EUAI}. Given the emergence of prompt-based AI systems --- that combine large pre-trained language models with dynamic prompting, retrieval-based augmentation, and adaptive context injection, as well as documentation artefacts --- prompt Cards may be essential to bridge the gap between regulatory intent, such as transparency and risk-governance, and real-world system complexity, notably evaluation and reproducibility.

\section*{Generative AI Usage Statement}
Generative AI (GPT-5.2, Gemini 3) was used to rectify grammar and spelling mistakes and for style editing.

\bibliographystyle{plain}
\bibliography{references}

\end{document}